\documentstyle[12pt,epsfig]{article}
\newcommand{\Nt}{\tilde{N}}
\def\lsim{\raise0.3ex\hbox{$<$\kern-0.75em\raise-1.1ex\hbox{$\sim$}}}
\def\gsim{\raise0.3ex\hbox{$>$\kern-0.75em\raise-1.1ex\hbox{$\sim$}}}
\begin{document}
\thispagestyle{empty}
 \mbox{} \hfill BI-TP 97/06\\
 \mbox{} \hfill ITP Budapest Rep. 529\\
 \mbox{} \hfill February 1997\\
\begin{center}
\vspace*{0.8cm}
{\huge \bf 
Screened Perturbation Theory }\\
\vspace*{1.0cm}
{\large  F. Karsch\\
Fakult\"at f\"ur Physik, Universit\"at Bielefeld,\\
D-33615, Bielefeld,
Germany \vspace{0.5cm}\\
A. Patk\'os and P. Petreczky\\
Department of Atomic Physics, E\"otv\"os University,\\
H-1088, Budapest,
Hungary} \\
\end{center}
\vspace*{1.0cm}
\begin{abstract}
A new perturbative scheme is proposed for the evaluation of
the free energy density of field theories at finite temperature.
The screened loop expansion takes into account exactly the phenomenon 
of screening in thermal propagators. The approach is tested 
in the $N$-component scalar field theory at 2-loop level and also at
3-loop in the large $N$ limit. The perturbative series generated
by the screened loop expansion shows much
better numerical convergence than previous expansions
generated in powers of the quartic coupling.
\end{abstract}
\vspace{1cm}
{\em PACS}: 11.10.Wx, 12.38.Mh\\
{\em Keywords:} QCD, scalar field theory,
free energy density, Debye screening, loop 
expansion,
gap equation, high-T expansion 

\newpage

\section{Introduction}
The weak coupling expansion of the QCD free energy density does show 
very bad 
convergence properties. The coefficients of the 
usual perturbative series are of alternating sign and of increasing 
magnitude 
\cite{Shu78}-\cite{Kas95}
Only in the TeV temperature 
range can one find a satisfactory numerical 
convergence rate \cite{Nie96}. This is
in great contrast to numerical calculations of this quantity (or of 
its 
derivative, the internal energy) in Monte Carlo simulations. These
calculations indicate that deviations from the high temperature ideal 
gas 
limit are within 15\% already for temperatures about twice the 
critical 
temperature \cite{Kar93}.

On the other 
hand it has been observed already quite some time ago that the 
formula for
the free energy density of a massive ideal gas does give a quite 
satisfactory
description of the numerical simulations \cite{Kar89}. Such a 
description leads
to results about 10\% below the Monte Carlo data. 
Although such an approach clearly has to be refined in order to take 
care
of the correct number of degrees of freedom and the modifications due 
to
interactions it indicates that a representation of a perturbative 
expansion
in terms of massive degrees of freedom might be a good starting 
point.

This observation leads us to conjecture that the apparent poor 
convergence of the 
perturbatively calculated free energy
density might be improved if the series is 
reorganised into a loop expansion evaluated with screened 
propagators;
{\it i.e.} instead of expanding around the massless ideal
gas limit, we intend to perform the loop expansion starting from a  
massive ideal gas. 

There are several screening scales in QCD, which makes the practical 
application of this conjecture more complicated, than it looks like 
at first sight. 
In the scalar theory, on the other hand, the only relevant mass is the 
Debye screening mass.
For this reason we will discuss here only  
the calculation of the free energy density of the $N$ component 
scalar $\Phi^4$-theory. The problem of poor numerical 
convergence shows up equally well already here 
\cite{Fren92,Par92,Par95}. 
One thus can test to what extent the above mentioned reorganisation 
of the perturbation series improves the convergence. 

We shall calculate and analyze first 
the two-loop correction to the massive 
ideal gas formula. With respect to their 
screening mass dependence our formulae
are exact. The ultra-violet finiteness of our calculation 
is ensured without the need to fix the screening mass 
inherently. As an external information we shall use for the screening 
mass
the root of the corresponding 1-loop gap equation 
\cite{Dol74,Buc94,Zwi93}. 
For large $N$ it also is possible to estimate the 3-loop correction 
of the
free energy density.
Finally, we shall compare
the results obtained with this {\em screened perturbative expansion} 
with
those obtained from the  conventional resummation scheme
\cite{Arn95,Fren92,Par92,Par95} as well as from an effective theory 
approach 
\cite{Bra95}.

\section{Loop Expansion with Debye Screened \hfill \break
Propa\-gators}
We consider an $O(N)$ symmetric scalar field theory with the 
following
Lagrangian:
\begin{eqnarray}
&
L=L_0+L_{int},\nonumber\\
&
L_0={1\over 2}({(\partial \phi_i)}^2+m^2 \phi^2_i),\nonumber\\
&
L_{int}=-{1\over 2} m^2 \phi_i^2+{g^2\over 24 N} {(\phi_i^2)}^2+
{g^2\over 24 N} (Z_2-1){(\phi_i^2)}^2.
\end{eqnarray}
Following ref. \cite{Fren92,Par92} we have introduced and substracted 
a mass term with $m\sim {\cal O} (g)$ in order to reorganize the 
perturbative 
expansion.
The coupling constant renormalisation factor is given by \cite{Amit}
\begin{equation}
Z_2=1+{3 g^2\over {(4 \pi)}^2} {N+8\over 18 \epsilon} + 
{\cal
O}(g^4),
\end{equation}
and we have left out the field renormalisation factor $Z_1$, which is 
unity
up to ${\cal O}(g^4)$.

All Feynman diagrams contributing to the free-energy density up to 
3-loops 
can be found in
\cite{Arn95}. The contribution of these diagrams are:
\begin{eqnarray}
-F^{(a)}&=&-{1\over 2} N J(m),\nonumber\\
-F^{(b)}&=&-{1\over 8} g^2 \Nt I(m)^2,\nonumber\\
-F^{(c)}&=&{1\over 2} m^2 N I(m),\nonumber\\
-F^{(d)}&=&-g^4 \Nt {N+8 \over 48 N} {1\over {(4 \pi)}^2 \epsilon} 
I(m)^2,
\nonumber\\
-F^{(e)}&=&{1\over 4} m^4 N I'(m)\nonumber,\\
-F^{(f)}&=&-{g^2 m^2\over 4} \Nt I(m) I'(m),\nonumber\\
-F^{(g)}&=&{g^4\over 16 N}  \Nt^2 I(m)^2 I'(m),\nonumber\\
-F^{(h)}&=&{g^4\over 48 N}  \Nt I_{ball}(m),\nonumber\\
F=F^{(a)}+F^{(b)}+F^{(c)}&+&F^{(d)}+F^{(e)}+F^{(f)}+F^{(g)}+F^{(h)}.
\label{f}
\end{eqnarray}
Here we have used the same labeling of the diagrams as in 
\cite{Arn95}. 
Furthermore, we use the abbreviation 
$\Nt=(N+2)/3$ and have introduced the 
following notation for the relevant integrals:
\begin{eqnarray}
J(m)&=&\int_P \log(P^2+m^2),\\
I(m)&=&\int_P {1\over P^2+m^2},\\
I'(m)&=&\int_P {1\over {(P^2+m^2)}^2},
\end{eqnarray}
\begin{equation}
I_{ball}={\displaystyle \int_{PQKR}{1\over 
{(P^2+m^2) (Q^2+m^2) (K^2+m^2) (R^2+m^2)}}}
\delta (P+Q+K+R).\nonumber\\
\end{equation}
Here we are using dimensional regularisation at finite
temperature. The symbol $\int_P$ 
therefore is understood to represent  
$T \sum_{n=-\infty}^{\infty} 
\int {d^{3-2 \epsilon}p\over {(2 \pi )}^{3-2 \epsilon}}$.
We note that all relevant sum-integrals appearing in Eqs.(3-7)
are UV-singular and do depend on the renormalisation scale $\mu$.
We shall argue below that a consistent expansion is obtained
by simply dropping these terms at each order in the loop expansion.

\noindent
At two loop level only the diagrams (a-c) contribute.
As will be discussed below we have, at present, difficulties to
extend our analysis at finite $N$ to 3-loops because we cannot
handle the finite terms in the ``basketball'' diagram (h) 
for nonzero values of the screening mass.
We have, however, also listed the 3-loop diagrams here, since 
the contribution from diagram (h) is suppressed in the large $N$
limit.  We are therefore also going to analyze the behaviour 
of our screened perturbative 
expansion in this limit by including the 3-loop contributions (d-g). 

We have explored two
possibilities to evaluate the 1-loop integrals $J(m)$, $I(m)$ and
$I'(m)$.
The first consists of performing the Matsubara sum using the contour
trick, subtract the $T=0$ piece, and then evaluate the remaining 
three-dimensional integral numerically\footnote{Before performing the
numerical evaluation of Eq.~(8) we have to expand
the integration measure in $\epsilon$.}.
For $J(m)$ we obtain
\begin{equation}
J(m)={\mu}^{2 \epsilon} \int {d^{3-2 \epsilon}p\over {(2 \pi )}^{3-2
\epsilon}}
\sqrt
{p^2+m^2}+ 2 {\mu}^{2 \epsilon} \int {d^{3-2 \epsilon}p\over {(2 \pi
)}^{3-2
\epsilon}}
\log(1-e^{-{\sqrt{p^2+m^2}\over T}}).
\end{equation}
The other two integrals, $I(m)$ and $I'(m)$, can be obtained from 
this by simple
differentiation with respect to $m^2$. The other possibility to 
evaluate these
sum-integrals is to use directly the high-T expansion
of the integrals $J(m)$, $I(m)$ and $I'(m)$ which can be
found in \cite{Arn93}.

The leading order 1-loop result is given by diagram (a), i.e. 
$F^{(a)}$.
It represents the contribution of a massive ideal gas. However,
on its own this contribution is not complete. It contains
UV-divergent and scale dependent contributions \cite{Arn93}. 
The divergent piece is proportional to $m^4 /\epsilon$.
Since the screening mass is linearly depending on $T$ this divergent
term cannot be canceled by the subtraction of appropriate zero 
temperature contributions. Higher loop contributions should take 
care of the cancellation of temperature dependent singularities. 
As the screening mass is ${\cal O} (gT)$ we expect that two and 
three loop contributions are needed to cancel the leading poles in
$\epsilon$. In fact, it is easy to check, that this is the case.
Collecting all 
poles in $\epsilon$ with residua proportional to 
$m^4,g^2T^2m^2,g^4T^4$
coming from diagrams up to 3-loop, we find that the singular terms
cancel irrespective of the actual value of $m^2$.

\vskip 5pt
This observation forms the basis for our conjecture: {\it The 
ultra-violet, temperature dependent singularities appearing in a 
given order of the screened loop expansion will be canceled by 
appropriate contributions from higher order loops. At any given
order in the expansion these singular contributions thus can be 
dropped.}
\vskip 5pt

At two loop order there appear also singularities with poles
in $\epsilon^2$ as well as poles with higher powers of $m^2$. 
In order to check explicitly the cancellation of these singularities
we would need to know the complete 4-loop contribution to the free
energy density. This is not known. We thus invoke the above 
conjecture to eliminate these singular terms from the expansion.

Another question which has to be investigated is the dependence on
the renormalisation scale. In the conventional perturbative approach
the scale dependence appears first at 3-loop level. The 
scale dependent contributions appearing in 
the 1-loop and 2-loop results (see
the explicit expressions of $J(m)$ and $I(m)$ in \cite{Arn93}) 
usually are canceled by corresponding zero temperature contributions.
In the screened perturbative approach this again cannot be achieved
because the scale dependent terms are proportional to the 
temperature,
i.e. the screening mass.
However, these additional scale  dependent terms, proportional to
the screening mass, are canceled by
higher loop contributions in analogy to the singular terms discussed
above. Furthermore,
the screening mass independent scale dependence of the 
3-loop contribution is exactly canceled when one takes into account
the scale dependence of the coupling constant in the 2-loop diagrams.
Up to ${\cal O}(g^4)$ the free energy is thus scale 
independent \cite{Bra95}.
We therefore omit terms proportional to $\ln({\bar \mu}^2/ T^2)$ 
in the lower loop expressions, 
too\footnote{$\bar \mu$ is 
the $\overline{\mbox{MS}}$ scale, $\mu^2={e^{\gamma_E}\over 4
\pi} {\bar \mu}^2$.}. 

Finally, we have to specify the screening mass.
Since $m^2$ has remained undetermined we have specified it as the exact 
root of the gap equation relevant for the chosen 
order of the loop expansion, i.e. we use the 1-loop gap 
equation\footnote{To make the 1-loop gap equation
self-consistent we have to substract singular and scale dependent terms
also here.
These terms would be generated by higher loop  contributions   (see \\
discussion in the next section ) which are 
not contained in the 1-loop gap equation (\ref{g}).} 
without expanding $I(m)$ in its argument:
\begin{equation}
m^2={g^2\Nt \over 2 N}(I(m)+{m^2\over 16 \pi^2}({1\over \epsilon}+
\ln{{\bar \mu}^2\over T^2})).
\label{g}
\end{equation}
Below we shall use roots of
Eq.~(\ref{g}) for the analysis of the free energy density. The screening
mass obtained this way agrees well with the 1-loop value for $g\lsim 1$
while it is only half as large as the 1-loop value for $g \sim 10$. 
We also note that as a consequence of using the root of the gap
equation for $m^2$, no term proportional to $g^2 m$ will appear in 
the free energy density. Such a term is seen to exist in $F^{(b)}$
when we use the high-T series for $I(m)$. However, this contribution 
is canceled by part of $F^{(c)}$, when exploiting (\ref{g}). The 
entire cubic dependence on $g^3$, which partly is responsible for 
the poor 
convergence of the conventional perturbative expansion, is now hidden in
a term proportional to $m^3$ which contributes to $J(m)$.

\section{Numerical Results}
We have evaluated the free energy density using Eq.~(\ref{f})
without making any assumption about the magnitude of $m$, i.e. we 
have
calculated $F$ "exactly" in $m$. For finite $N$ we  
can do this so far only at 2-loop level. At
3-loop one would have to calculate the finite part of $I_{ball}(m)$ 
for 
arbitrary values of $m$. Since we are not assuming that $m$ is
small we did use the integral representations for $J(m)$, $I(m)$
and $I'(m)$ in our numerical calculations. However, we also checked
the alternative approach based on the high-T expansion and found that 
this series does converge rapidly. 
For the integrals listed above it is sufficient to take the first
six terms of high-T expansion for $g \in [0, 12]$.
	
In Fig 1. we show the 1-loop and 2-loop results for the free energy 
density as function of the self-coupling $g$ and for $N=1$. The 
screening
mass has been computed from Eq. (\ref{g}). Also shown there are
the results obtained with the conventional perturbative expansion up 
to
${\cal O}(g^2)$ and ${\cal O}(g^3)$ in the scalar self-coupling. 
The Figure clearly shows that the difference between 1-loop and 
2-loop
results of the screened perturbative expansion
is small and essentially stays constant over the entire range of
couplings. In the conventional perturbative approach on the other hand
different orders in the expansion differ largely from each other.

For large $N$ the contribution of the basketball diagram is 
suppressed
by $1/N$ relative to other diagrams. In this limit we thus can 
consider
also the 3-loop result.
In the large-$N$ limit the
1-loop gap equation (\ref{g})  
becomes exact. Its derivation as a Schwinger-Dyson equation has been
given in \cite{Dol74}. The setting sun diagram with exact vertex
becomes negligible in the large N limit, while the divergent and scale
dependent part of the tadpole contribution is canceled by the
renormalisation of the scalar self-coupling.

For our numerical investigation of the large-N limit we have chosen $N=10$. 
The 1-loop, 2-loop and 3-loop results for the
free energy density obtained from  the screened perturbative
expansion are shown in Fig.~2. The screening mass has been determined
from Eq.~(\ref{g}). For comparison
we also show the conventional ${\cal O}(g^4)$ result of \cite{Arn95}.
Lower orders in this approach do show alternating behaviour similar to that 
shown in Fig.~1 for the case $N=1$.
We note that like in the $N=1$ case the higher loop corrections lead
to smaller deviations from the massless ideal gas than the leading
1-loop result.
We also find that for $g<12$ the
difference between the 3-loop result and the massless ideal gas limit 
is only $10\%$.

\section{Conclusions}

We have formulated a perturbative scheme for the $N$-component 
scalar field theory which includes a temperature
dependent screening mass in the scalar propagator. We conjecture
that a systematic loop expansion is obtained when singular terms
and scale dependent terms proportional to the screening mass are 
dropped order by order in the perturbative expansion. This expansion
coincides to all orders with the conventional approach, if the screening
mass is treated perturbatively. Within the screened perturbative 
approach we can handle the effect of a screening mass exactly.

The analysis of the $N=1$ scalar field theory and its large-$N$ limit
to 2-loop and 3-loop, respectively, shows that within this approach 
the
convergence of the perturbative series is drastically improved over
the conventional approach. It is also evident that taking into account
higher orders in the massive loop expansion systematically brings the 
free energy density closer to the massless ideal gas limit. 
It certainly would be interesting to test these results against 
non-perturbative lattice calculations.

We hope that the nice properties seen here for the scalar theory 
remain valid also in the case of QCD where more mass scales will
become relevant. 

\vskip 0.5truecm
\noindent
{\bf Acknowledgement}
\vskip 0.3truecm
P.P. thanks ~J. Pol{\'o}nyi and Z. Fodor for useful discussions.

\newpage
\vspace{1cm}
{\Large \bf Figure Caption}
\begin{figure}[htp]
\epsfig{file=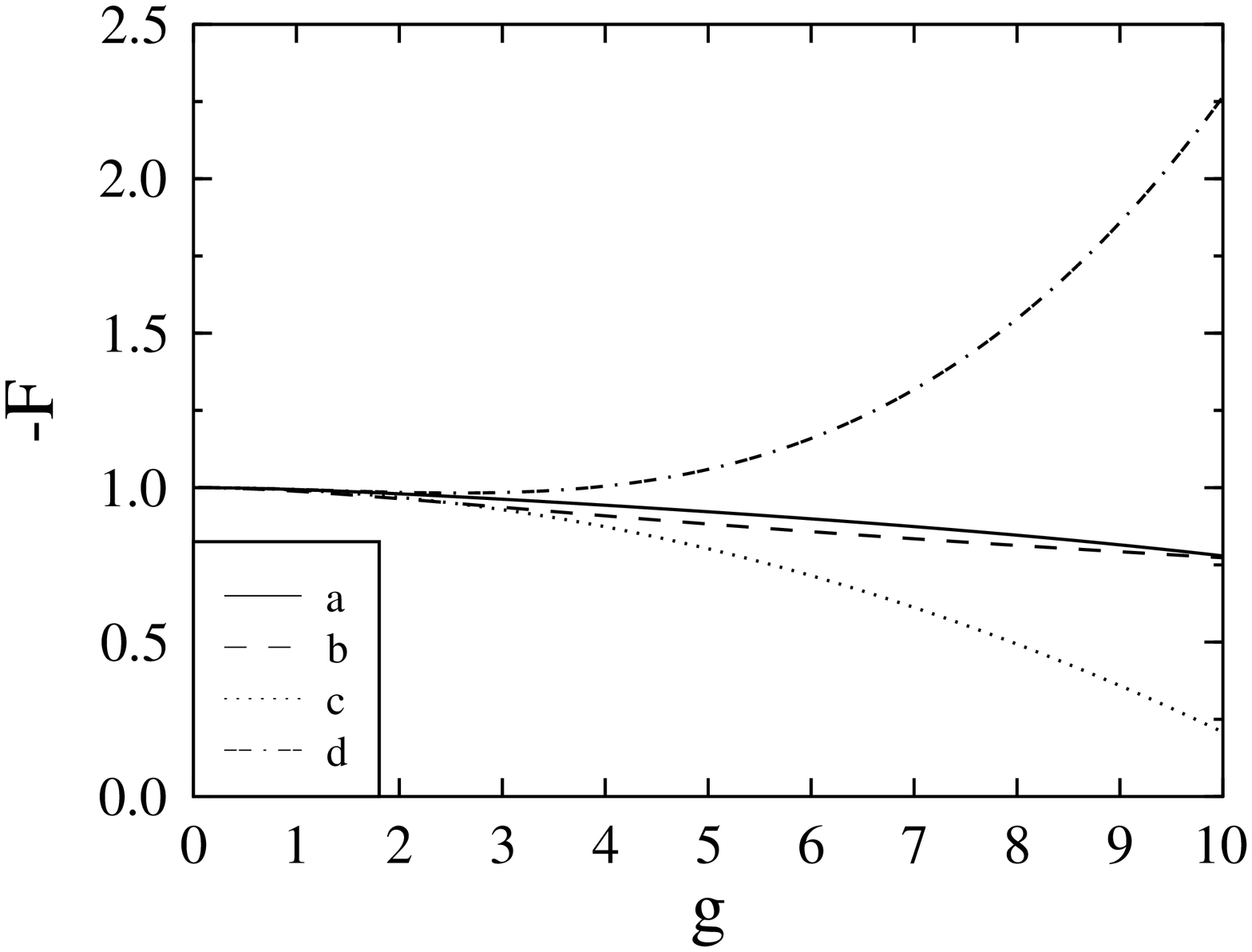,height=9truecm}
\caption{\em Shown is the free-energy density of the N=1 scalar 
field theory 
as a function of the scalar self-coupling g in the units of the 
free energy density of a massless ideal gas ($\pi^2 T^4\over 90$). 
The curves
represent 2-loop (a) and 1-loop (b) results of the screened loop 
expansion as well as the 
${\cal O}(g^2)$ (c) and  ${\cal O}(g^3)$ (d) results of the 
conventional perturbative expansion.}
\label{fig:1}
\end{figure}
\vspace{-3.4cm}
\begin{figure}[t]
\epsfig{file=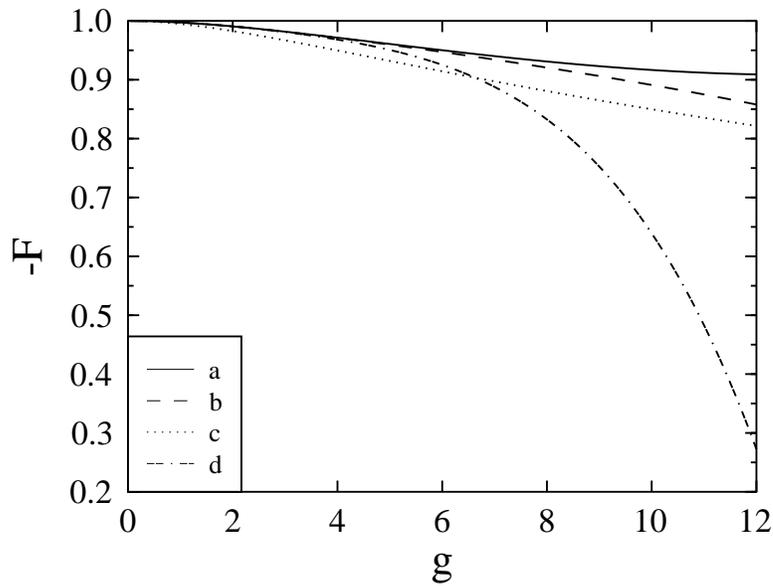,height=9truecm}
\caption{\em The free energy density of the scalar field theory in
the large-$N$ limit as
a function of the scalar self-coupling $g$ in units of the free 
energy
density of a massless ideal gas. Shown are results for $N=10$ (see 
text).
The curves represent 3-loop (a) 2-loop (b) and 1-loop (c) results of 
the screened loop expansion as well as the 
${\cal O} (g^4)$ (d) results of the conventional 
perturbative expansion.}
\label{fig:2}
\end{figure}

\end{document}